\documentclass[12pt]{article}
\usepackage{latexsym} \usepackage{epsf}
\usepackage{a4}
\usepackage{amsfonts}
\newcommand{\oB}{\vert_{\partial M}=0} 
\newcommand{\be}{\begin{equation}}
\newcommand{\ee}{\end{equation}}
\newcommand{\bea}{\begin{eqnarray}}
\newcommand{\eea}{\end{eqnarray}}
\newcommand{\beq}{\begin{eqnarray}}
\newcommand{\eeq}{\end{eqnarray}}
\newcommand{\beao}{\begin{eqnarray*}}
\newcommand{\eeao}{\end{eqnarray*}}

\newcommand{\pa}{\partial}
\newcommand{\e}{{\rm e}}

\renewcommand{\d}{{\rm d}}

\newcommand{\Ref}[1]{(\ref{#1})}

\newcommand{\bc}{boundary conditions }

\newcommand{\gse}{ground state energy }

\newcommand{\E}{{\cal E}}

\begin{document}
\title{Casimir force between Chern-Simons surfaces} 
\author{
{\sc M. Bordag}\thanks{e-mail: Michael.Bordag@itp.uni-leipzig.de} 
~~ and ~ 
{\sc D.V. Vassilevich}\thanks{On leave from Department of Theoretical Physics,
     St.Petersburg University,
     198904 St.Petersburg, e-mail: Dmitri.Vassilevich@itp.uni-leipzig.de}\\
\small  University of Leipzig, Institute for Theoretical Physics\\
\small  Augustusplatz 10/11, 04109 Leipzig, Germany
}\date{hep-th/9911179}
\maketitle
\begin{abstract}
We calculate the Casimir force between two parallel plates
if the boundary conditions for the photons are modified due
to presence of the Chern-Simons term. We show that this
effect should be measurable within the present experimental
technique.
\end{abstract}
\section{Introduction}\label{Sec1}
The Chern-Simons gauge theories \cite{Deser:1982vy,Schonfeld:1981kb}
attract much attention due to both their theoretical beauty and
practical applications to certain condensed matter phenomena. Most
notable is the generation of states with fractional statistics first
observed by Wilczek \cite{Wilczek:1982du}.  Later this phenomenon was
used to describe composite fermions \cite{Heinonen} in the theory of
the fractional quantum Hall effect (FQHE) (see, e.g., 
\cite{Das-Sarma-Pinczuk}).

The Casimir effect (see, e.g., the book \cite{mostepanenkotrunov}) is
a nowadays also experimentally well established
\cite{Roy:1999dx,Lamoreaux:1997wh} macroscopic quantum effect. Several
applications are known, ranging from the bag model in QCD
\cite{Hasenfratz:1978dt} to constraints on hypothetical long range
interactions \cite{Bordag:1999gx}.

In the present paper we investigate another situation where the
Casimir effect eventually may serve as a test for certain theoretical
models in solid state theory, namely for models including a
Chern-Simons term.  We consider the situation when the Maxwell volume
action (which is in (3+1) dimensions) is supplemented by a
Chern-Simons surface term. We argue that such boundary term naturally
leads to a modification of ordinary conductor and bag boundary
conditions. Such a modification can also be considered on its own
right as the only possible one with a local gauge invariant $P$-odd
term without any new dimensional coupling.  The boundary condition and
quantization are discussed in the next section.

In section 3 we calculate the ground state energy for the system of two
parallel planes bearing different Chern-Simons charges. We show that the
Casimir force depends substantially on these charges and can even change sign
under certain conditions.  Given recent advances in the Casimir force
experiments \cite{Lamoreaux:1997wh}, the effect could be measurable.
Theoretical and phenomenological implications of our work are discussed in the
last section.

\section{Chern-Simons boundary conditions for ordinary
and dual potentials}\label{Sec2}
Consider a region $M$ in the space-time with a boundary
$\partial M$. Let the Maxwell action on $M$ be supplemented
by the Chern-Simons surface action:
\begin{equation}
S=-\frac 14 \int_M d^4x\ g^{\frac 12} F_{\mu\nu}F^{\mu\nu} 
-\frac a2 \int_{\partial M} d^3x \ \varepsilon^{ijk} A_i 
\partial_j A_k ,
\label{action}
\end{equation}
where $\varepsilon^{ijk}$ is the Levi-Civita tensor and the $x^j$ 
are coordinates on the boundary $\partial M$, $a$ being a real 
parameter. After integration by parts one gets
\begin{eqnarray} 
S&=&-\frac 12 \int_M d^4x\ g^{\frac 12} 
A_\mu (-g^{\mu\nu}\Box +\nabla^\nu \nabla^\mu )A_\nu \nonumber \\ 
\ &\ & -\frac 12 \int_{\partial M} d^3x 
(h^{\frac 12}(\partial_NA_i-\partial_iA_N)A^i+ a\varepsilon^{ijk} A_i 
\partial_j A_k) , 
\label{Stot} 
\end{eqnarray} 
where $\Box$ is the d'Alambertian, $N$ is the outward pointing 
normal vector, and $h$ is the determinant of the induced metric 
on $\partial M$. The volume term in (\ref{Stot}) (with a suitable
gauge choice) generates the wave equation for $A_\mu$, the
surface term generates the boundary conditions. 

There are two
sets of local gauge-invariant boundary conditions which
ensure the vanishing of the surface term in (\ref{Stot}). The first
one is called relative boundary conditions 
\begin{equation}
A_i\oB , \qquad (\partial_N+k)A_N\oB.
\label{relbc}
\end{equation} 
Here $k$ is the trace of the second fundamental form of the boundary.
The second set reads \cite{Elizalde:1998ha}:
\begin{equation}
A_N\oB ,\quad 
(\partial_NA_i+a{\varepsilon_i}^{jk}\partial_jA_k) 
\oB \ \ i=0,1,2, . 
\label{bcon} 
\end{equation} 
Note that variation of the surface term in (\ref{Stot})
with respect to $A^i$ gives precisely the sum of the two
conditions (\ref{bcon}).

Let us suppose for simplicity that the boundary $\partial M$
consists of two parallel infinite plates located at $x^3=const.$
In this case $k=0$. Consider the right plate, $x^N=x^3$. The
boundary conditions (\ref{relbc}) and (\ref{bcon}) can be
rewritten in terms of the field strengths giving
respectively
\begin{eqnarray}
&&H_3\oB ,\qquad E_\parallel \oB ;\label{relstr}\\
&&(E_3+aH_3)\oB ,\qquad (H_\parallel -aE_\parallel )\oB
\label{absstr}
\end{eqnarray}

When the coupling to charged particles is absent (as in the
problem of the Casimir energy calculations) one can quantize
the photons in terms of the dual potentials $A^*_\mu$:
\begin{equation}
{}^*F^{\mu\nu}=\frac 12 \varepsilon^{\mu\nu\rho\sigma}
F_{\rho\sigma}, \qquad 
{}^*F_{\mu\nu}=\partial_\mu A^*_\nu -\partial_\nu A^*_\mu
\label{dualpot}
\end{equation}
In the case of the electromagnetic field in a dielectric the
quantization in terms of $A^*$ even gives some technical
advantages \cite{Bordag:1999vs}. 
One can also consider the Chern-Simons term with the ordinary vector
potential $A_\mu$ replaced by the dual potential $A^*_\mu$.
The corresponding boundary conditions will be dual to the two sets
(\ref{relstr}) and (\ref{absstr}):
\begin{eqnarray}
&&E_3\oB ,\qquad H_\parallel \oB ;\label{drelstr}\\
&&(H_3-aE_3)\oB ,\qquad (E_\parallel +aH_\parallel )\oB
\label{dabsstr}
\end{eqnarray}

The conditions (\ref{relstr}) correspond a conducting boundary.
Their duals (\ref{drelstr}) are called bag boundary conditions.
The conditions (\ref{absstr}) and (\ref{dabsstr}) generalize
(\ref{drelstr}) and (\ref{relstr}) respectively for the case
of non-zero Chern-Simons coupling. The two sets (\ref{absstr})
and (\ref{dabsstr}) are related by $a\to -1/a$. Therefore,
it is enough to consider just one set (\ref{absstr}). 
Different connected pieces of the boundary can correspond
to different values of $a$. Equal values of $a$ everywhere
will correspond to complete dual invariance.

At this point we must note that there is no exact equivalence
between boundary conditions and a boundary action. Here we are
considering a ``mixed'' problem when the $a=0$ physics
is described by rigid boundary conditions which are later
modified by the presence of the Chern-Simons boundary term.
A more rigorous way would be to describe the full interaction of
photons with the boundary by a boundary action and treat it
as a $\delta$-function potential (for the
mathematical set-up
see e.g. \cite{Bordag:1999ed}). This would lead however to considerable
technical complexities. We believe that our present approach
gives correct (zeroth order) approximation to the real
physics.

A suitable gauge choice is
\begin{equation}
\partial^j A_j=0 \,.\label{gauge}
\end{equation}
The boundary conditions (\ref{bcon}) are invariant under the
gauge transformations $A\to A+\partial \omega$ provided the
gauge parameter $\omega$ satisfies 
\begin{equation}
\partial_N \omega \oB \ . \label{ghost}
\end{equation}
Upon quantization (\ref{ghost}) become boundary conditions
for the ghost field. Propagators for the ghosts and for
the $A_3$ does not depend on the Chern-Simons coupling $a$.
Moreover, in the gauge (\ref{gauge}) they does not contain
$\partial_3$ and, hence, does not contribute to the Casimir
force. Therefore, in the following we will be interested in
the two polarizations of the field $A_i$ satisfying (\ref{gauge}). 

By performing the Fourier transformation in the $x^j$ directions
we arrive at the problem of diagonalization of the operator
${L_i}^k ={\varepsilon_i}^{jk}k_j$. With our sign conventions
${\varepsilon_0}^{12}=1$.
The following two
polarizations solve the problem
\begin{equation}
A^\pm_i=\pmatrix{k_a^2 \cr 
k_1k_0 \mp k_2 \sqrt{k_a^2-k_0^2} \cr
k_2k_0 \pm k_1 \sqrt{k_a^2-k_0^2} } b^\pm (x^3) \,,
\end{equation}
$i=\{ 0,a\}$, $a=1,2$.
Both vectors $A^\pm$ satisfy the gauge condition (\ref{gauge}).
They correspond to the eigenvalues $\pm \sqrt{k_a^2-k_0^2}$
of the operator ${L_i}^k$.

\section{Casimir force}\label{Sec3}
We define the ground state energy density per unit area
of the plates in the usual way by
\be\label{gse} 
\E_{0}=\frac12\sum_{n}\int{\d k_{a}\over (2\pi)^{2}}\left(k_{a}^{2}+
\omega_{n}^2\right)^{\frac12-s} \,,
\ee
where $\omega_n$ is discrete momentum in the third direction,
$s>\frac32$ is the (zeta-functional) regularization parameter with $s\to
0$ in the end. 

Consider two planes located at $x^3=0,L$ with different Chern-Simons
charges, $a_{(0)}$ and $a_{(L)}$. From the equation of motion
we note $k_{0}^{2}=k_{a}^{2}+k_{3}^{2}$. The boundary conditions
on the two planes for the modes $b^\pm$ become:
\be\label{bc} 
\left({\pa\over\pa x^{3}}\pm a_{(0,L)}
  k_3 \right)b^{\pm}(x^{3})_{\vert_{x^{3}=0,L}}=0 
\ee
Note that $\partial_N\vert_{x^3=L}=-\partial_N\vert_{x^3=0}=\partial_3$.
The sign before the Levi-Civita tensor should be also different on the
two components of the boundary due to the reversed orientation.
Consider the $b^+$ polarization for definiteness. The boundary condition
at $x^3=0$ is solved by the following harmonics:
\be\label{pol} 
b^{+}(x^{3})=-a_{(0)}\sin k_{3}x^{3} +\cos k_3x^3\, .
\ee
Substituting (\ref{pol}) in the boundary condition at $x^3=L$ one gets
\be\label{freqcond}
(a_{(L)}-a_{(0)})\cos k_{3}L-(1+a_{(L)}a_{(0)})\sin k_{3}L=0 \,,
\ee
whose solutions are $k_{3}=\omega_{n}$. It is useful to rewrite condition
\Ref{freqcond} in the form $f(k_{3})=0$ with
\be\label{fc2}
f(k)= \sin(k L+\delta)
\ee
where $\delta=\arctan \left((a_{(0)}-a_{(L)})/(1+ a_{(L)}a_{(0)})\right)$. 
Obviously, when $\delta=0$ ($a_{(0)}=a_{(L)}$) the Chern-Simons
surfaces interact with the same potential as two Dirichlet planes.
In the case 
$\delta=\frac{\pi}2$ interaction coincides with that of a Dirichlet with
a Newmann plane.

To proceed with the calculation of the \gse we first integrate in \Ref{gse}
over the momenta $k_{a}$ and obtain
\be\label{gse1} 
\E_{0}={-1\over 12\pi}{1\over 1-\frac32 s} \zeta_{\rm cs}(s)\,,
\ee
where we introduced the 'Chern-Simons' zeta function
\be\label{zcs} 
 \zeta_{\rm cs}(s)=\sum_{n}\omega_{n}^{3-2s}\,.
\ee
Now we proceed in the usual way by changing from the discrete sum  
to an integral. Following, e.g., the paper \cite{Bordag:1995jz} we obtain
\be\label{zcs2} \zeta_{\rm cs}(s)=\int_{\gamma}{\d k\over 2\pi
  i}k^{3-2s}{\pa\over\pa k} \ln f(k)\,, \ee
where the integration path $\gamma$ encircles the zeros of the function
$f(k)$. Having in mind that $\delta=0$ corresponds to the well known case of
Dirichlet \bc on both planes we represent the \gse in the form
\be\label{gse2}
\E_{0}=-{1\over L^{3}}{\pi^{2}\over 1440} h(\delta) \,,
\ee
where the function $h(\delta)$ describes the relative deviation of the
Casimir energy for the 
Chern-Simons boundary conditions from that for two 
Dirichlet planes. From \Ref{gse1} and
\Ref{zcs2} we obtain 
\be\label{h1}
h(\delta)={120\over\pi^{2}} \ \zeta(s\to 0) 
\ee
where $s\to 0$ means the analytic continuation and we can put $L=1$ in
$\zeta$.

\begin{figure}[!htbp]\unitlength=1cm
\begin{picture}(5,8)
\put(0,0){\epsfxsize=12cm 
\epsffile{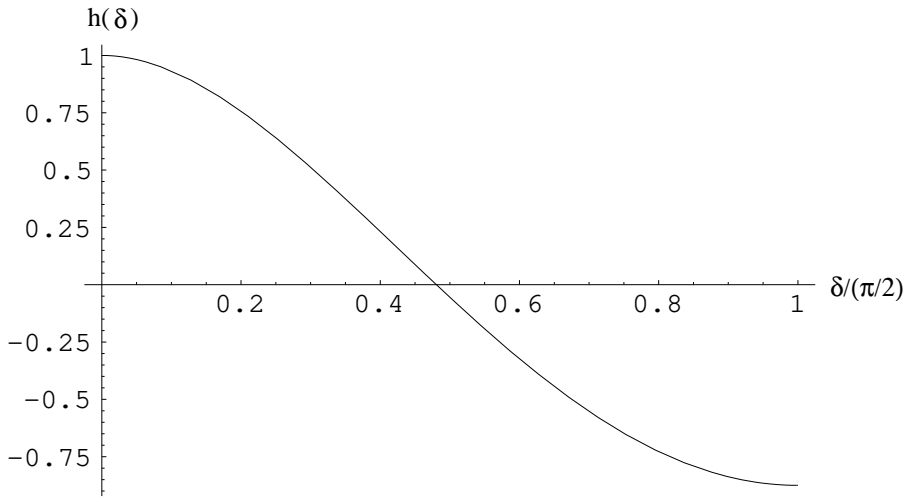}}
\end{picture}
\caption{The function $h(\delta)$ showing the dependence of the \gse on the
  Chern-Simons parameter 
$\delta=\arctan \left((a_{(0)}-a_{(L)})/(1+ a_{(L)}a_{(0)})\right)$ .  }
\label{figure1}
\end{figure}

In order to calculate the function $h(\delta)$ we divide the integration path
$\gamma$ in \Ref{zcs2} into an upper part $\gamma_{1}$ (with $\Im k>0$) and an
lower part $\gamma_{2}$. On the upper part we represent 
\be\label{upper} {\pa\over\pa k} \ln \sin(k+\delta)={\pa\over\pa
  k}\left(-i(k+\delta)+\ln\left(1-\e^{2i(k+\delta)}\right)\right) \ee
and on the lower part we choose \be\label{lower} {\pa\over\pa k} \ln
\sin(k+\delta)={\pa\over\pa k}\left(i(k+\delta)
  +\ln\left(\e^{-2i(k+\delta)}-1\right)\right) \,, \ee which is just the
complex conjugate of (\ref{upper}). The first contributions on the r.h.s. of
(\ref{upper}) and (\ref{lower}) do not depend on the Chern-Simons parameter
$a$ and represent (divergent) energy density in the empty Minkowski space.  We
drop them therefore. Than we turn the integration contours, namely $k\to ik$
on $\gamma_{1}$ and $k\to -ik$ on $\gamma_{2}$. After that we can take the
limit $s\to 0$ because of the now exponential convergence of the integral and
obtain
\be\label{zcs3} \zeta = \int_{0}^{\infty}{\d k\over 2\pi} \ k^{3}{\pa\over\pa
  k}\ln \left(1+\e^{-4k}-2\cos(2\delta)\e^{-2k}\right)\,.  \ee
For the function $h(\delta)$ we obtain after some trivial transformations
\be\label{h2} h(\delta)={-120\over \pi^{3}} \ \int_{0}^{\infty} \d k \ k^{3} \ 
{\e^{-2k}-\cos (2\delta)\over \cosh (2k)-\cos(2\delta)} \,.  \ee
We note the special values $h(0)=1$ which confirms the case of Dirichlet \bc
on both planes and
\be\label{spv}
h({\pi\over4})=-\frac7{128} ~~ \mbox{\rm ,} ~~ h({\pi\over2})=-\frac78 \, , 
\ee
where the last number reproduces the repulsive potential between Dirichlet and
Neumann planes.  This function can be easily plotted, the result is shown in
Figure \ref{figure1}. To obtain the complete interaction energy one should also
take into account the contribution of the second polarization $b^-$ which
results in an overall factor of 2 in the ground state energy (\ref{gse1}).
Obviously, this means that the full interaction energy of the Chern-Simons
surfaces is obtained from the interaction energy of two conducting planes by
multiplication by the same function $h(\delta )$.

\section{Conclusions}\label{Sec4}
In this paper we have argued that the presence of the Chern-Simons boundary
action naturally leads to a modification of the boundary conditions for the
photons. More precisely, we show that one of the admissible sets of the
boundary conditions, namely ``bag'' or ``conductor'' ones, receives a
contribution from the Chern-Simons interaction (depending on which potentials,
ordinary or dual, are used in the Chern-Simons term). Usually, the type of the
potential does not matter because the Chern-Simons term originates from the
dual invariant volume action $a\int_M {}^*F^{\mu\nu}F_{\mu\nu}$.  This case
corresponds to equal Chern-Simons couplings on both surfaces and does not lead
to modification of the ground state energy. In the present context the case of
the broken duality ($a_{(0)}\ne a_{(L)}$) is more interesting, but less clear
as far as a modification of the boundary conditions is concerned.  To clarify
this point one should consider the microscopic interaction of photons with the
surface which is out of the scope of the present paper (see also a remark
before the eq. (\ref{gauge})).  In any case, we find it quite unnatural if the
presence of a specific $P$-odd surface interaction will have no effect on the
boundary conditions at all. The simplest relevant modification of the boundary
conditions (local, gauge invariant, without new dimensional parameters) is
just the one considered in the present paper.

For a given set of the Chern-Simons boundary conditions we have calculated
rigorously the ground state energy for two parallel planes bearing different
Chern-Simons charges. As can be seen from the fig. 1, the Casimir force
exhibits a strong dependence on the Chern-Simons coupling and can even change
its sign as compared to the case of two conducting planes. With the present
experimental technique \cite{Roy:1999dx} maintaining a precision of about 1\%
even correspondingly small values of the Chern-Simons parameter $a$ should be
measurable in an experiment with one conducting plane and another one
presumably bearing the Chern-Simons interaction.  Such a measurement would
provide an important check for the theoretical models including $P$-odd
interaction of electromagnetic field with a surface.

The most spectacular application would be the (FQHE). An exact relation
between the mechanism of the fractional conductivity and the boundary
conditions is not well established yet. As well, it is not clear which part of
the virtual photons responsible for the Casimir force would ``see'' the FQHE.
However, if any interaction exists and if it makes sense to idealize it by
boundary conditions than it should be of the type considered here. Hence, the
Casimir force between two surfaces might serve as a tool to study these
effects. Even though an actual experimental realization of such a measurement
is not going to be simple, it could result in independent and interesting
results.

\section*{Acknowledgments}
This work has been supported in part by the Alexander von Humboldt
foundation and by RFBR, grant 97-01-01186.

\bibliographystyle{unsrt}
\bibliography{chs}
\end{document}